\documentclass[aps, superscriptaddress, reprint]{revtex4-1}

\usepackage{lipsum}
\usepackage{graphicx}
\usepackage{soul}
\usepackage{xfrac}
\usepackage[colorlinks,
	linkcolor=blue,
	urlcolor=blue,
	citecolor=blue]{hyperref}
\usepackage[T1]{fontenc}       % Use modern font encodings
\usepackage[version=3]{mhchem}
\usepackage{xcolor}
\usepackage{titlesec}
\usepackage{palatino} % font to avoid the ugly default µ
\titlespacing*{\section}{0pt}{1.1\baselineskip}{\baselineskip}
\usepackage{pgffor}
\usepackage{pdfpages}
\usepackage{etoolbox} 
\makeatletter
\patchcmd{\@outputpage@head}{\@ifx{\LS@rot\@undefined}{}{\LS@rot}}{}{}{}
\makeatother

\begin{document}
	\title[]{Balanced electron--hole transport in spin-orbit semimetal \ce{SrIrO3} heterostructures}
	\author{Nicola \surname{Manca}}
	\email{nicola.manca@spin.cnr.it}
%	\thanks{\\Present Address: Physics Department, University of Genova, Via Dodecaneso 33, 16146 Genova, Italy}
	\affiliation{Kavli Institute of Nanoscience, Delft University of Technology, P.O. Box 5046, 2600 GA Delft, The Netherlands}

	\author{Dirk J. \surname{Groenendijk}}
	\affiliation{Kavli Institute of Nanoscience, Delft University of Technology, P.O. Box 5046, 2600 GA Delft, The Netherlands}

	\author{Ilaria \surname{Pallecchi}}
	\affiliation{CNR-SPIN, c/o Dipartimento di Fisica, University of Genoa, Via Dodecaneso 33, 16146 Genova, Italy 3Consiglio}
		
	\author{Carmine \surname{Autieri}}
	\affiliation{Consiglio Nazionale delle Ricerche CNR-SPIN, UOS L'Aquila, Sede Temporanea di Chieti, 66100 Chieti, Italy}

	\author{Lucas M. K. \surname{Tang}}
	\affiliation{High Field Magnet Laboratory (HFML - EMFL), Radboud University, Toernooiveld 7, Nijmegen, The Netherlands}

	\author{Francesca \surname{Telesio}}
	\affiliation{Dipartimento di Fisica, University of Genoa, Via Dodecaneso 33, 16146 Genova, Italy}
		
	\author{Giordano \surname{Mattoni}}
	\affiliation{Kavli Institute of Nanoscience, Delft University of Technology, P.O. Box 5046, 2600 GA Delft, The Netherlands}		

	\author{Alix \surname{McCollam}}
	\affiliation{High Field Magnet Laboratory (HFML - EMFL), Radboud University, Toernooiveld 7, Nijmegen, The Netherlands}

	\author{Silvia \surname{Picozzi}}
	\affiliation{Consiglio Nazionale delle Ricerche CNR-SPIN, UOS L'Aquila, Sede Temporanea di Chieti, 66100 Chieti, Italy}
		
	\author{Andrea D. \surname{Caviglia}}
	\affiliation{Kavli Institute of Nanoscience, Delft University of Technology, P.O. Box 5046, 2600 GA Delft, The Netherlands}
	
	\date{\today}
	
	\begin{abstract}
		Relating the band structure of correlated semimetals to their transport properties is a complex and often open issue. The partial occupation of numerous electron and hole bands can result in properties that are seemingly in contrast with one another, complicating the extraction of the transport coefficients of different bands. The $5d$ oxide \ce{SrIrO3} hosts parabolic bands of heavy holes and light electrons in gapped Dirac cones due to the interplay between electron-electron interactions and spin-orbit coupling. We present a multifold approach relying on different experimental techniques and theoretical calculations to disentangle its complex electronic properties. By combining magnetotransport and thermoelectric measurements in a field-effect geometry with first-principles calculations, we quantitatively determine the transport coefficients of different conduction channels. Despite their different dispersion relationships, electrons and holes are found to have strikingly similar transport coefficients, yielding a holelike response under field-effect and thermoelectric measurements and a linear, electronlike Hall effect up to 33~T.
	\end{abstract}
		
	\maketitle

Oxide heterostructures have been intensely studied in recent years as a versatile platform for controlling electronic properties of materials~\cite{Zubko2011}. Charge transfer, strain engineering, and polar instabilities are part of the toolbox available at oxide interfaces for controlling phases of matter, such as two-dimensional superconductors and various magnetic ground states. Although the manipulation of broken symmetries is a well-developed topic in the field, experimental control of topological phases at oxide interfaces has so far been elusive. Symmetry-protected boundary states in oxide heterostructures have been considered theoretically as a promising route to realize novel topological materials. Much attention has been focused on \ce{SrIrO3} and its heterostructures as candidates for correlated topological insulators~\cite{Pesin2010, Kargarian2011}, topological semimetals~\cite{Yang2014a, Chen2015a}, topological Hall effect~\cite{Matsuno2015} and unconventional superconductors~\cite{Wang2011, Kim2012c, Meng2014}. In its bulk form this material exhibits a nodal line at the $U$ point and characteristic transport signatures of Dirac electrons, such as large and linear magnetoresistances~\cite{Carter2012,Zeb2012,Fujioka2017}. However, when synthesised as an epitaxial thin film, \ce{SrIrO3} shows transport characteristics that are inconsistent with this picture and not yet understood, including a linear and strongly reduced magnetoresistance~\cite{Biswas2014,Zhang2015c}. Angle-resolved photoemission spectroscopy shows that the degeneracy at the Dirac points is lifted, leading to a Fermi surface with pockets of light electrons together with heavy holes~\cite{Nie2015,Liu2016a}. X-ray diffraction studies have correlated this modification of the electronic structure to an epitaxially stabilised lattice distortion that breaks the orthorhombic bulk symmetry~\cite{Liu2016,Horak2017}. An understanding of the charge and transport properties of this correlated semimetal is a fundamental step for the realization of topological phases in oxide heterostructures and is developed here.

Here, we report on an extensive characterization of the transport properties of heteroepitaxial \ce{SrIrO3 / SrTiO3} by combined field-effect, magnetotransport and thermoelectric measurements. Numerical analysis, supported by first principles calculations, account for the coexistence of an electron-like Hall effect with a hole-like electrical conductivity and thermopower. The emerging picture of a compensated semimetallic state harmonizes transport and spectroscopic data.

\ce{SrIrO3} thin films were grown by pulsed laser deposition on single-crystal \ce{SrTiO3 [001]} substrates, and then encapsulated \textit{in-situ} with a \ce{SrTiO3} layer to prevent degradation during lithographic processing~\cite{Groenendijk2016b}.
Details of growth conditions and sample characterization of \ce{SrIrO3 / SrTiO3} heterostructures are discussed in Ref.~\onlinecite{Groenendijk2016b}. 
Magnetotransport measurements were acquired in a four-probe configuration in a flow cryostat with a base temperature of 1.6~K.
The Seebeck effect was measured in a physical properties measurement system by Quantum Design equipped with thermal transport option in the continuous scanning mode with a 0.4~K/min cooling rate. 
First-principles density functional theory calculations were performed within the generalized gradient approximation (GGA) by using the plane wave \textsc{vasp}~\cite{Kresse1999} package and the PBEsol for the exchange-correlation functional~\cite{Perdew2008} with spin-orbit coupling.
The Hubbard $U$ effects on the \ce{Ir} sites were included within the GGA$+U$~\cite{Anisimov1991} approach using the rotational invariant scheme~\cite{Liechtenstein1995}. With $U$ larger than 1~eV, the bulk is magnetic. To deal with the non magnetic bulk \ce{Ir} compounds~\cite{Kim2017}, we assumed $U=0.80~\mathrm{eV}$ and $J_H=0.15~U$.
The core and the valence electrons were treated with the projector augmented wave method~\cite{Blochl1994} and a cutoff of 400~eV for the plane wave basis was used. An 8$\times$8$\times$6 $k$-point Monkhorst-Pack grid~\cite{Pack1977} was used for the calculation of the bulk phase.

\begin{figure}
	\includegraphics[width=\linewidth]{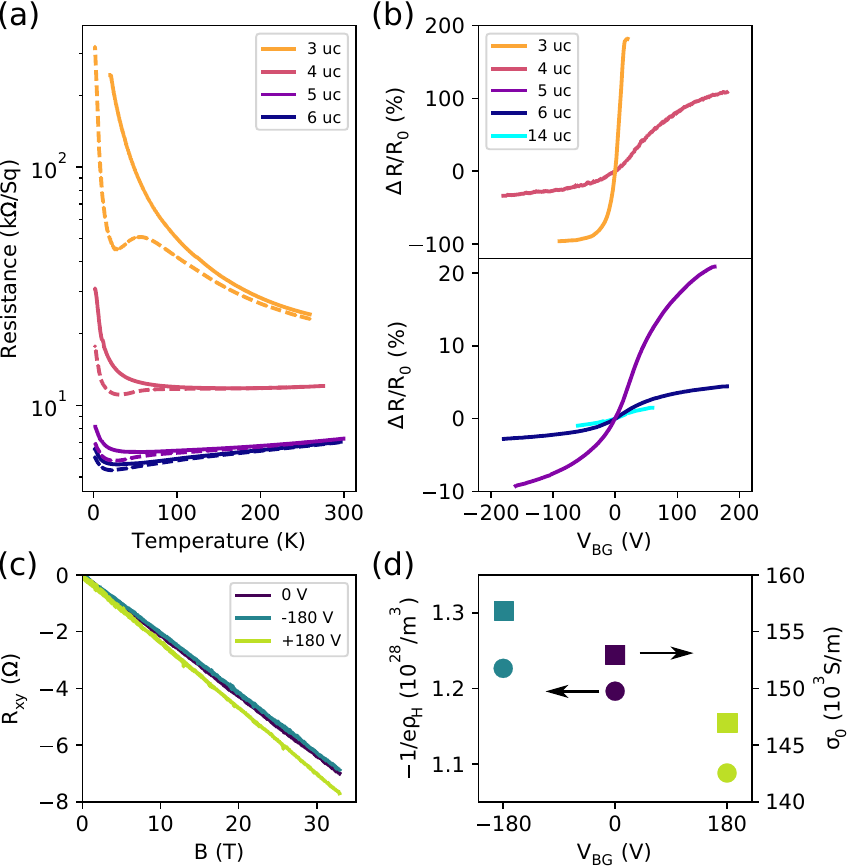}
	\caption{\label{fig:DC_transport}
		Magnetotransport under field effect in \ce{SrIrO3} thin films.
		(a) $R(T)$ for different thicknesses measured at $V_{\mathrm{BG}}=0$~V (the solid lines) and $V_{\mathrm{BG}}=-180$~V (the dashed lines).
		(b) Relative variation of the electrical resistance at 1.6~K vs $V_{\mathrm{BG}}$, both metallic and semiconducting samples show the same qualitative response.
		(c) Hall resistance of a six unit cell (uc) thin film for different $V_{\mathrm{BG}}$'s at 1.6~K.
		(d) Free electron density calculated from (d) in a single band picture (circles) and $\sigma_0(V_{\mathrm{BG}})$ (the squares). 
	}		
\end{figure}

Figure~\ref{fig:DC_transport}(a) shows resistance vs temperature [$R(T)$] characteristics of \ce{SrIrO3} films of different thicknesses (solid lines).
%As recently reported, a metal-insulator transition occurs between 4 and 3 unit cells (uc)~\cite{Groenendijk2017, Schutz2017}.
As recently reported, a metal-insulator transition occurs between four and three ucs and bulklike electrical resistivity is reached above six ucs~\cite{Groenendijk2017, Schutz2017}.
Through the application of a back-gate voltage ($V_{\mathrm{BG}}$), the carrier density of the \ce{SrIrO3} films can be tuned by field-effect. This technique is often used in low-density two-dimensional systems or semiconductors to change the carrier density and consequently the resistance. The dashed $R(T)$ curves in Fig.~\ref{fig:DC_transport}(a) show a huge resistance change upon the application of $V_{\mathrm{BG}}=-180$~V, indicative of the low carrier density of this system. The field effect efficiency decreases with the thickness and above six ucs the $R(T)$ plots measured with $V_{\mathrm{BG}}=0$~V and $V_{\mathrm{BG}}=-180$~V are not distinguishable. 
The way the electric field affects the $R(T)$ characteristics gives us a first hint of the carrier type of the system:
A negative $V_{\mathrm{BG}}$ lowers the resistance as expected from a conductor whose carriers are holes. 
Figure~\ref{fig:DC_transport}(b) shows the relative variation of the electrical resistance, measured on samples of different thicknesses, while sweeping $V_{\mathrm{BG}}$ at 1.6~K.
The hole-like response is consistent over the whole range of thicknesses explored, independent of the semiconducting or metallic behavior~\cite{Groenendijk2017}.
Reducing the \ce{SrIrO3} thickness the gating efficiency becomes more pronounced because of the decreased screening effect from the free carriers, and for the three ucs case we can even reach an insulating state by the field effect [Fig.~\ref{fig:DC_transport}(b)].
At base temperature it is possible to observe a field effect even for thicknesses above six ucs and in Fig.~\ref{fig:DC_transport}(b) we show the 14~uc case (cyan plot) where, as expected from the strong screening effect, the signal is very small, on the order of a few percent.
We note that the response to the back gate is mediated by the dielectric constant of the \ce{SrTiO3} substrate ($\varepsilon_r^{STO}$). 
The nonlinear temperature and electric field dependence of $\varepsilon_r^{STO}$ determine the non monotonic behavior of the $R(T)$'s with applied $V_{\mathrm{BG}}$ in Fig.~\ref{fig:DC_transport}(a) and the reduced gating efficiency at high voltages observed in Fig.~\ref{fig:DC_transport}(b)~\cite{Neville1972,Davidovikj2017}.
In the Supplemental Material, Sec.~I we provide further evidence of a hole-dominated electrical conductivity ($\sigma_0$), showing the effect of doping \ce{SrIrO3} thin films with oxygen vacancies~\footnote{See Supplemental Material at [URL will be inserted by publisher] for the effect of oxygen vacancies doping, details of the sampling algorithm and the analysis of the Magneto-Resistance.}.

Previous literature reports showed that the Hall effect in \ce{SrIrO3} thin films is negative and almost linear~\cite{Zhang2015c}, which is at odds with the hole-type field-effect response. To study this, we choose a thickness of six ucs, which is small enough to be tunable by a field effect and large enough to be sufficiently conductive at low temperatures. 
The corresponding Hall effect measured at 1.6~K is presented in Fig.~\ref{fig:DC_transport}(c).
It is linear and negative up to 33~T, similar to what would be observed in a system dominated by a single band of electrons.
However, the response of the Hall signal ($\rho_\mathrm{H}$) and conductivity at zero magnetic field ($\sigma_0$) to the $V_\mathrm{BG}$, reported in Fig.~\ref{fig:DC_transport}(d), show that such a simple picture is inadequate.
Despite its negative slope, the Hall signal responds to the back-gate as if the electrical transport is dominated by hole carriers.
Furthermore, the carrier density calculated in a single band picture ($1/\rho_\mathrm{H} \approx 10^{28}~m^{-3}$) would make the back gate almost ineffective because of the strong screening effect.
The discrepancy between Hall and field-effect data is a clear indication of the multi-band character of this system.

The Hall resistivity of two parallel channels of holes and electrons is given by:
\begin{equation}
\label{eq:Hall_effect_standard}
\rho_{\rm{H}} \equiv \frac{t R_{xy}}{B} = \frac{1}{e}\frac{n_{\mathrm{h}} \mu_{\mathrm{h}}^2-n_{\mathrm{e}} \mu_{\mathrm{e}}^2 + (n_{\mathrm{h}}-n_{\mathrm{e}})(\mu_{\mathrm{h}} \mu_{\mathrm{e}} B)^2}{\sigma_0^2/e^2 + (n_{\mathrm{h}}-n_{\mathrm{e}})^2(\mu_{\mathrm{h}} \mu_{\mathrm{e}} B)^2},
\end{equation}
where
\begin{equation}
\label{eq:Sigma_0}
\sigma_0 = \sigma_{\mathrm{h}}+\sigma_{\mathrm{e}} = e(n_{\mathrm{h}} \mu_{\mathrm{h}}+ n_{\mathrm{e}} \mu_{\mathrm{e}}),
\end{equation}
and $t$ is the film thickness,  $R_{xy}$ is the Hall resistance, $e$ is the elementary charge, $B$ is the magnetic field, $n$ is the carrier density, $\mu$ is the mobility, and $\mathrm{e}$ and $\mathrm{h}$ indicate electrons and holes, respectively.
Since the measured Hall effect from Fig.~\ref{fig:DC_transport}(c) is linear and negative we can approximate Eq.~\ref{eq:Hall_effect_standard} with its low-field limit,
\begin{equation}
\begin{aligned}
\label{eq:Hall_effect_approx}
\rho_{\rm{H}} = \frac{e}{\sigma_0^2} \left ( n_{\mathrm{h}} \mu_{\mathrm{h}}^2-n_{\mathrm{e}} \mu_{\mathrm{e}}^2 \right).
\end{aligned}
\end{equation}
Standard analysis on two-band systems is based on the combined measurement of $\sigma_0$ and $\rho_{xy}$ where the presence of flex points in the magnetic field dependence of $\rho_{xy}$ provides a powerful constraint for the calculation of carrier densities and mobilities of the conduction channels. 
Because in our case such flex points are likely out of the probed range ($\pm33$~T), the extraction of the carrier parameters in \ce{SrIrO3} thin films remains undetermined.

\begin{figure}[t]
	\includegraphics[width=\linewidth]{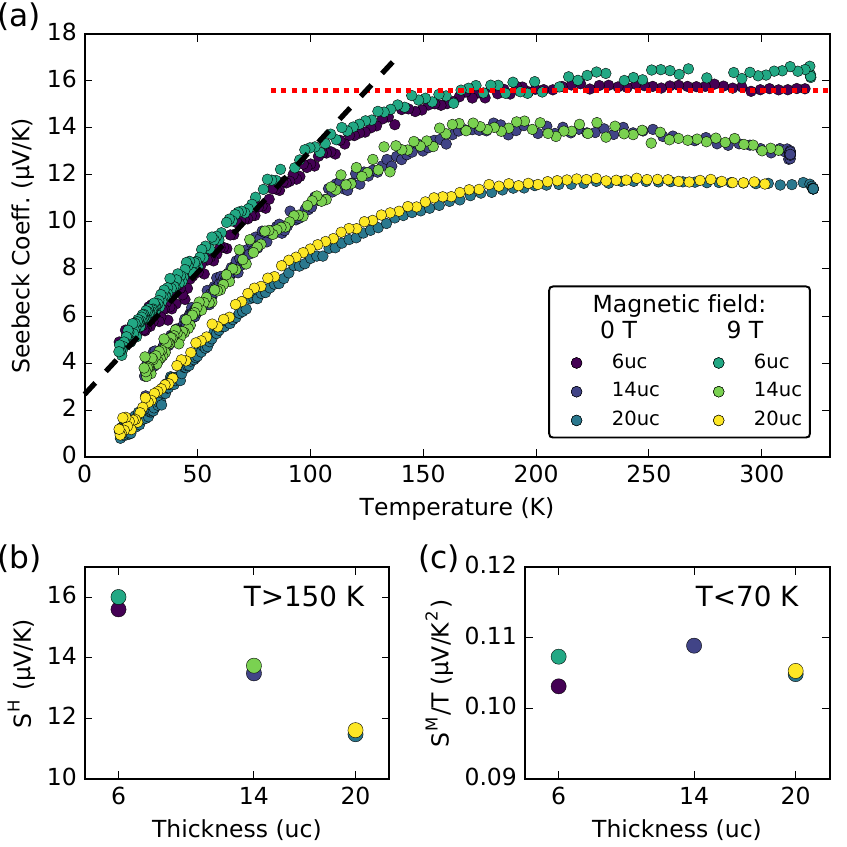}
	\caption{\label{fig:Seebeck}
		Thermoelectric measurements on \ce{SrIrO3} thin films.
		(a) Seebeck coefficient as a function of temperature, thickness and magnetic field. 
		(b) Seebeck coefficient corresponding to Eq.~\ref{eq:Seebeck_Heikes} and
		(c) Seebeck slope used in eq.~\ref{eq:Seebeck_Mott}.
		The dashed lines in (a) show a linear fit used to extract the data in (b) (red dotted line) and (c) (black dashed line), corresponding to the high ($>$150~K) and and low ($<$70~K) temperature regime.
	}
\end{figure}

Thermoelectric measurements can provide complementary information to magnetotransport, allowing us to identify the transport coefficients of the two carriers.
For this experiment we prepared a dedicated series of samples having size of $10\times5~mm^{2}$ and thicknesses of six, 14 and 20 ucs. Samples thinner than six ucs were not measurable because of the high noise at the electrical contacts, in particular at low temperatures.
We thus measured the Seebeck coefficient ($S$) of metallic \ce{SrIrO3} films as a function of thickness, temperature, and magnetic field.
The experimental results are presented in Fig.~\ref{fig:Seebeck}(a), where all the samples show a nearly constant positive value $\mathrm{S\approx10\!-\!20~\mu V/K}$ above 150~K and a linear decrease below 70~K.
As opposed to what is observed in bulk \ce{SrIrO3} samples~\cite{Pallecchi2016}, here the Seebeck coefficient does not show multiple sign changes, indicating a drastically different electronic structure. A further difference is the absence of any magnetic-field dependence of the thermoelectric response, indicating that in thin films the Seebeck effect is dominated by the diffusive mechanism.
The high ($>$150~K )and low ($<$70~K) temperature regimes are well described by the formulas of Heikes~\cite{Heikes1964} and Mott~\cite{Jonson1980}, respectively, which provide a direct relationship between the measured quantities and the microscopic material properties. For each single band of either holes or electrons, the formulas of Heikes ($S\mathrm{^H}$) and Mott ($S\mathrm{^M}$) Seebeck coefficients are as follows:
\begin{align}
\label{eq:Seebeck_Heikes}
S\mathrm{^H_{h/e}} &= \pm \frac{k_{\rm{B}}}{e} \log\left( \frac{2-n\nu}{n\nu} \right), \\
\label{eq:Seebeck_Mott}
S\mathrm{^M_{h/e}} &= \pm \left( \frac{3}{2}-\alpha \right)
\left( \frac{2 (2\pi)^{8}}{3^{5}} \right)^{1/3}
\frac{k_{\rm{B}}^2}{e \hbar^2}
\frac{m^*}{n^{2/3}} T,
\end{align}

\begin{figure}[t]
	\includegraphics[width=\linewidth]{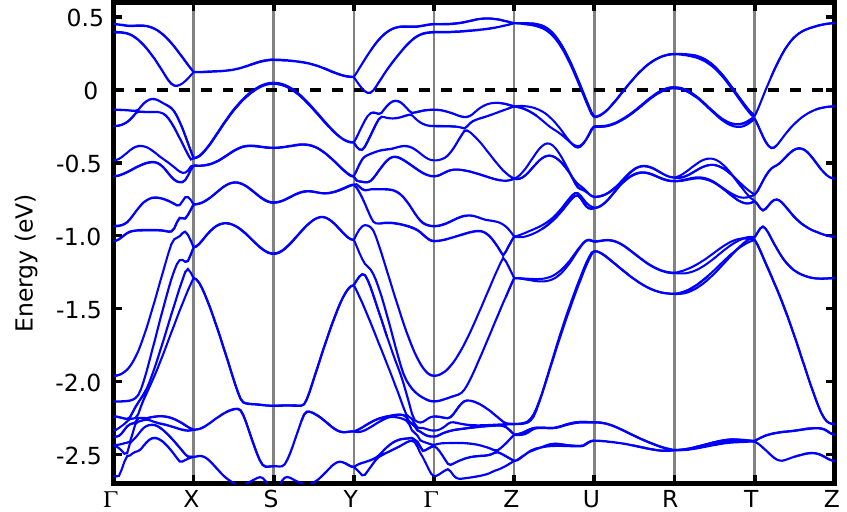}
	\caption{\label{fig:BS}
		Electronic band structure of bulk \ce{SrIrO3} at $U$=$0.80$~eV with lattice constants $a$=3.905~\AA~and $c$=4.08~\AA.
	}
\end{figure}

\begin{figure*}
	{
		\includegraphics[width=0.75\linewidth]{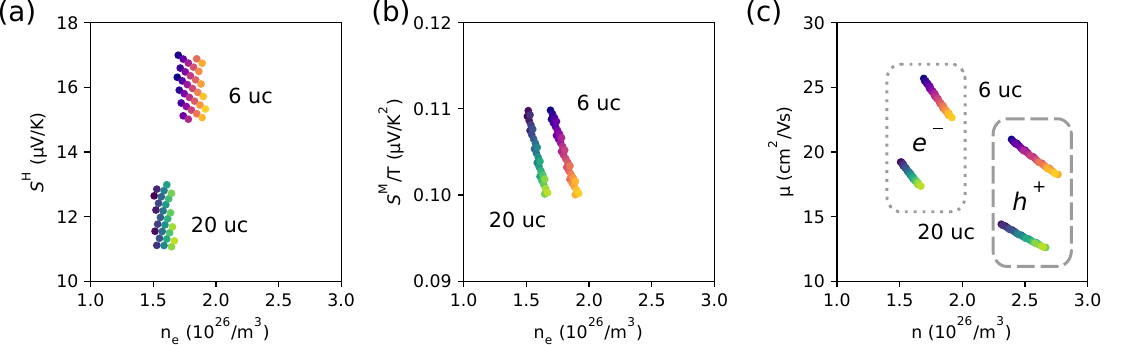}
		\caption{\label{fig:Simulation}
			Transport coefficients of 20 and 6~uc \ce{SrIrO3} thin films.
			(a) Total Seebeck coefficients from the formulas of Heikes and (b) Mott of acceptable $\{n_i,\mu_i\}$ combinations. Here, only electrons are shown for clarity.
			(c) Free-carrier densities and mobilities for electrons (points) and holes (the dashed line) satisfying the experimental constraints. 
		}
	}
\end{figure*}
\noindent
where $k_{\rm{B}}$ is the Boltzmann constant, $\hbar$ is the reduced Planck constant, $\nu$ is the unit cell volume, $\alpha$ is a parameter related to the scattering mechanisms ($0\leq\alpha\leq1$), $m^*$ is the effective mass of the carriers and the $\pm$ sign corresponds to holes or electrons, respectively.
In a multiband picture the total Seebeck coefficient in diffusive regime can be expressed as 
\begin{equation}
S = \frac{e}{\sigma_0} (n_{\mathrm{h}} \mu_{\mathrm{h}} |S_{\mathrm{h}}| - n_{\mathrm{e}} \mu_{\mathrm{e}} |S_{\mathrm{e}}|).
\end{equation}
That in combination with Eqs.~\ref{eq:Seebeck_Heikes} and \ref{eq:Seebeck_Mott} allows us to write the Seebeck coefficient in the two temperature regimes,
\begin{align}
\label{eq:Seebeck_Heikes_carr}
\mathrm{S^H} &= \frac{k_{\mathrm{B}}}{ \sigma_0} \left(
n_{\mathrm{h}} \mu_{\mathrm{h}} \log \left (\frac{2-n_{\mathrm{h}} \nu}{n_{\mathrm{h}} \nu} \right) -
n_{\mathrm{e}} \mu_{\mathrm{e}} \log \left (\frac{2-n_{\mathrm{e}} \nu}{n_{\mathrm{e}} \nu} \right)
\right), \\
\label{eq:Seebeck_Mott_carr}
\mathrm{S^M}  &= \frac{ e}{ \sigma_0} \xi \left( n_{\mathrm{h}}^{1/3} \mu_{\mathrm{h}} m_{\mathrm{h}}^* - n_{\mathrm{e}}^{1/3} \mu_{\mathrm{e}} m_{\mathrm{e}}^* \right) T,
\end{align}
where $\xi$ is the numerical pre-factor appearing in Eq.~\ref{eq:Seebeck_Mott},
$$
\xi = \left( \frac{3}{2}-\alpha \right)
\left( \frac{2 (2\pi)^{8}}{3^{5}} \right)^{1/3}
\frac{k_{\rm{B}}^2}{e \hbar^2}
$$
The experimental values for $S\mathrm{^H}$ and $S\mathrm{^M}/T$ are presented in Figs.~\ref{fig:Seebeck}(b) and (c) as a function of the film thickness, where $S\mathrm{^H}$ is calculated as the average of $S$ above 150~K and $S\mathrm{^M}/T$ is determined from linear fitting of $S$ below 70~K.
Although the thermoelectric response at low temperature $S\mathrm{^M}/T$ does not show any thickness dependence, at high temperature $S\mathrm{^H}$ decreases linearly with the thickness. This could be an indication of a different balance of electrons and holes when approaching the metal-insulator transition~\cite{Groenendijk2017}.

The analysis of the thermoelectric response  relies on fixing the values of the free parameters of Eq.~\ref{eq:Seebeck_Mott_carr} ($\alpha,m^*_{\mathrm{e}},m^*_{\mathrm{h}}$) on the basis of theoretical considerations.
Here, $\alpha=0.5$ was used, which is the typical choice when scattering is dominated by impurities or phonons~\cite{Pallecchi2010}.
To estimate the effective masses in \ce{SrIrO3} thin films, we study the electronic structure of \ce{SrIrO3} in the thick-film limit by means of first principles calculations. The in-plane lattice parameters was fixed to the value of the \ce{SrTiO3} substrate (3.905~\AA), and the out-of-plane lattice parameter was fixed to the experimental value for \ce{SrIrO3} (4.08~\AA)~\cite{Groenendijk2016b}.
Figure~\ref{fig:BS} shows the corresponding band structure, where the density of states near the Fermi level is dominated by the $5d$ $t_{2g}$ contribution as in orthorhombic bulk \ce{SrIrO3}.
We calculate the effective masses for the holes at the maxima of the dispersion relations close to the Fermi level located at the $S$ and $R$ points, and find an average value of $m^*_{\mathrm{h}}=1.55~m_{\mathrm{e}}$.
At the $U$ and $T$ points and along the $Y$-$\Gamma$ directions, we have the minima, and the corresponding average effective mass for the electrons is $m^*_{\mathrm{e}}=0.34~m_{\mathrm{e}}$.
These values are in agreement with the effective masses extracted from angle-resolved photoemission spectroscopy measurements performed on \ce{SrIrO3 / SrTiO3} heterostructures~\cite{Liu2016a}.
\begin{table}[b]
	\begin{center}
		\begin{tabular}{c  c  c  c  c }
			\hline\hline
			$t$ (uc)
			& ~$\rho_{\mathrm{H}}\,(\mathrm{n\Omega\, m / T})$ 
			& ~ $\sigma_0\,(\mathrm{{S/m}})$
			& ~ $\mathrm{S^H}\,(\mathrm{\mu V/K})$
			& ~ $\mathrm{S^M}\,(\mathrm{nV/K^2})$\\[1.4ex]
			\hline
			30 & $-1.27$ & $1.0\cdot10^5$ & $12 \pm 1$ & $105 \pm 5$\\[1.1ex]
			6 & $-0.45$ & $1.5\cdot10^5$ & $16 \pm 1 $ & $105 \pm 5$\\[1.1ex]
			\hline\hline
		\end{tabular}
	\end{center}
	\caption{Input parameters of the sampling algorithm.}
	\label{tab:Parameters}
\end{table}
\begin{table}[b]
	\begin{center}
		\begin{tabular}{c  c  c  c  c  c  c}
			\hline\hline
			t
			& ~$n_{\mathrm{e}}$
			& ~$n_{\mathrm{h}}$
			& ~$\mu_{\mathrm{e}}$
			& ~$\mu_{\mathrm{h}}$
			& ~$\sigma_{\mathrm{e}}$
			& ~$\sigma_{\mathrm{h}}$\\
			 ~(uc)
			& ~$(\sfrac{10^{20}}{cm^3})$
			& ~$(\sfrac{10^{20}}{cm^3})$
			& ~$(\sfrac{cm^2}{V\,s})$
			& ~$(\sfrac{cm^2}{V\,s})$
			& ~$(\sfrac{S}{m})$
			& ~$(\sfrac{S}{m})$\\[1.4ex]
			\hline
			20 
			& 1.6${\times}10^6$
			& 2.5${\times}10^6$
			& 18
			& 13.5 
			& 4.6
			& 5.4 \\[1.1ex]
			6 
			& 1.8${\times}10^6$
			& 2.6${\times}10^6$
			& 27
			& 22
			& 7.8
			& 9.1 \\[1.1ex]
			\hline\hline
		\end{tabular}
	\end{center}
	\caption{Charge carriers characteristics extracted from the sampling analysis with the experimental constraints.}
	\label{tab:Results}
\end{table}
In the following, we evaluate $n_{\mathrm{e}},\ \mu_{\mathrm{e}},\ \mathrm{and}\ n_{\mathrm{h}},\ \mu_{\mathrm{h}}$ by using a direct sampling algorithm.
Equations \ref{eq:Sigma_0}, \ref{eq:Hall_effect_approx}, \ref{eq:Seebeck_Heikes_carr} and \ref{eq:Seebeck_Mott_carr} are linearly independent and can therefore be combined to determine $n_i$ and $\mu_i$.
For each combination of $(n_{\mathrm{e}}, \mu_{\mathrm{e}})$ the corresponding $(n_{\mathrm{h}}, \mu_{\mathrm{h}})$ pair is calculated by using the experimental $\sigma_0$ and $\rho_{\rm{H}}$ in Eqs.~\ref{eq:Sigma_0} and \ref{eq:Hall_effect_approx}.
The resulting $(n_i,\mu_i)$ set is accepted if the Hall effect calculated with Eq.~\ref{eq:Hall_effect_standard} and the thermoelectric coefficients $S^{\rm{H}}$ and $S^{\rm{M}}/T$, calculated with Eqs.~\ref{eq:Seebeck_Heikes_carr} and \ref{eq:Seebeck_Mott_carr}, agree with the experimental data (see also the Supplemental Material, Sec.~II for further details~\cite{Note1}).
Table \ref{tab:Parameters} shows the experimental values used as input parameters, where the $\pm$ on the Seebeck coefficients indicates the range of the acceptance condition. Since the values of $\rho_\mathrm{H}$ and $\sigma_0$ above 15~ucs show no thickness dependence~\cite{Groenendijk2017}, we combine electrical transport data from a thick (30-uc) sample with the 20-uc Seebeck data to perform the analysis.
The calculated $(n_i,\mu_i)$ combinations for both 20 and 6~ucs are presented in Figs.~\ref{fig:Simulation}(a) and \ref{fig:Simulation}(b), whereas Fig.~\ref{fig:Simulation}(c) shows the corresponding transport coefficients.
Each $(n_i,\mu_i)$ set is a possible solution satisfying the the experimental constraints, and multiple sets are accepted because of the tolerances reported in Table \ref{tab:Parameters}.
We find that the carrier density and mobility for electrons and holes must be located in two closely spaced groups of points on the $(n,\mu)$ plane, with the electrons having higher mobility and lower density than the holes.
Table \ref{tab:Results} reports the centers of these groups together with the corresponding conductance of each channel.
The ratio between electron and hole carrier densities is not dramatic ($\approx1.7$), confirming the marked multi-band character of this system.
The results from this analysis are consistent with the experimental observation of a hole-dominated electrical conductivity ($\sigma_{\mathrm{h}} > \sigma_{\mathrm{e}}$), although this constraint was not explicitly  introduced into the analysis. This is in agreement with both back-gate experiments presented in Fig.~\ref{fig:DC_transport} and the oxygen vacancies doping experiment reported in the Supplemental Material, Sec.~I~\cite{Note1}.
The calculated $n_{\mathrm{e}}$'s are two orders of magnitude lower than that obtained in a single band picture, showing that the measured Hall signals are determined by carrier compensation.
These values have a weak temperature dependence, since both $\rho_{\mathrm{H}}$ and $\sigma_0$ show small temperature variations~\cite{Groenendijk2017} and the Seebeck coefficients at low and high temperatures are in good agreement with Eqs.~\ref{eq:Seebeck_Heikes} and \ref{eq:Seebeck_Mott}.
From the results of Table~\ref{tab:Results} it is possible to calculate the cyclotron component of the magnetoresistance and compare it with the measured one. This is discussed in the Supplemental Material, Sec.~III~\cite{Note1} where we show that our analysis is compatible with the experimental data in the framework of the present literature~\cite{Abrikosov1998,Parish2003,Parish2005,Gerber2007,Hu2008a,Huynh2011,Kozlova2012,Alekseev2015}.  

In conclusion, we investigated the electronic structure of \ce{SrIrO3} thin films by means of multiple transport techniques. The semimetallic nature of \ce{SrIrO3} manifests itself in a Hall effect dominated by electrons, and a hole-like electrical conductivity and Seebeck effect.
The combination of magnetotransport and thermoelectric measurements with first-principles calculations allows for obtaining a limited ensemble of possible transport coefficients for the charge carriers.
Our results indicates that electrons and holes have similar densities and mobilities, yet the higher conductivity of the hole channel makes it dominant in the electrical transport.
This analysis constitutes a comprehensive and robust description of the electronic structure of \ce{SrIrO3}, paving the way for future studies on \ce{SrIrO3}-based heterostructures
and that can be extended to unravel the electronic structure of other semimetallic compounds.

\section*{Acknowledgements}

This Rapid Communication was supported by The Netherlands Organisation for Scientific Research (NWO/OCW) as part of the Frontiers of Nanoscience program (NanoFront), by the Dutch Foundation for Fundamental Research on Matter (FOM) and by CNR-SPIN via the Seed Project ``CAMEO''. 

\bibliographystyle{apsrev4-1}
\bibliography{library}
\newpage\newpage



\section*{Supplementary material}

# Balanced electron–hole transport in spin-orbit semimetal SrIrO$_3$ heterostructures

Nicola Manca,[1,*] Dirk J. Groenendijk,[1] Ilaria Pallecchi,[2] Carmine Autieri,[3] Lucas M. K. Tang,[4] Francesca Telesio,[5] Giordano Mattoni,[1] Alix McCollam,[4] Silvia Picozzi,[3] and Andrea D. Caviglia[1]

[1]*Kavli Institute of Nanoscience, Delft University of Technology,*
*P.O. Box 5046, 2600 GA Delft, The Netherlands*

[2]*CNR-SPIN, c/o Dipartimento di Fisica, University of Genoa,*
*Via Dodecaneso 33, 16146 Genova, Italy 3Consiglio*

[3]*Consiglio Nazionale delle Ricerche CNR-SPIN,*
*UOS L'Aquila, Sede Temporanea di Chieti, 66100 Chieti, Italy*

[4]*High Field Magnet Laboratory (HFML - EMFL),*
*Radboud University, Toernooiveld 7, Nijmegen, The Netherlands*

[5]*Dipartimento di Fisica, University of Genoa,*
*Via Dodecaneso 33, 16146 Genova, Italy*

(Dated: February 16, 2018)

**Sec. 1. Doping by oxygen vacancies**

Oxygen vacancies ($V_O$) are an ubiquitous electron-doping mechanism for oxide materials and can be reversibly induced by controlling the oxygen partial pressure and the sample temperature. $V_O$ act as electron donors, resulting in an enhanced or reduced conductivity depending on whether electrons or holes are the dominant carriers in the electrical transport.

In order to verify that the DC electrical conductivity is dominated by holes, the effect of doping by oxygen vacancies ($V_O$) in a $SrIrO_3$ thin film is reported in Fig. S1. Here, a thickness of 3 uc was chosen to enhance the gas exchange, thanks to the high surface/volume ratio. Oxygen vacancies were created by increasing the sample temperature in an environment with low oxygen partial pressure. We measured the sample resistance in a four probe geometry in a temperature- and pressure-controlled sample holder. The electrical resistance of the sample was monitored as a function of temperature and oxygen partial pressure following the sequence illustrated in Fig. S1(a), where the green curve was

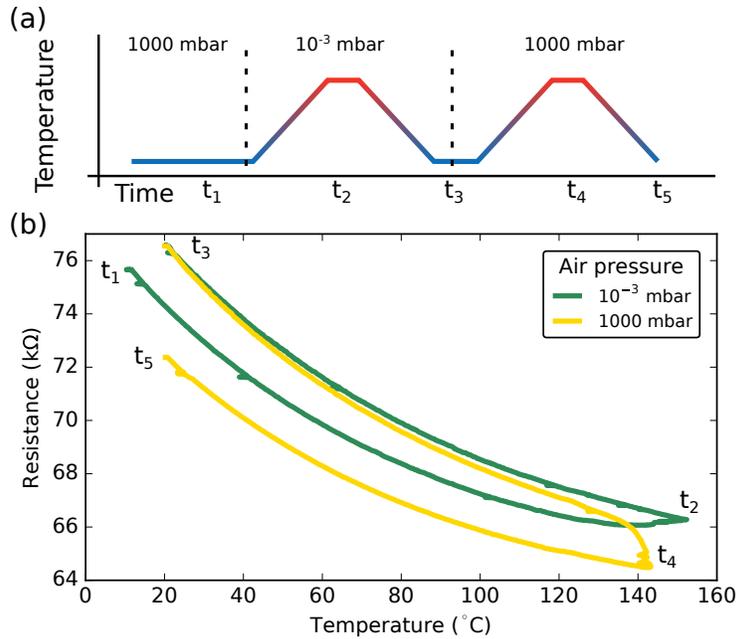

Figure S1. Doping of a 3 uc $SrIrO_3$ thin film by oxygen vacancies. (a) Schematic representation of a sequence of temperature ramps between 20 °C and 150 °C under different air pressure conditions. Ramp speed is 1 K/s. (b) Resistance vs temperature curves corresponding to the labels $t_i$. At high temperature and low pressure $V_O$ are created and the electrical resistance increases.

measured in $10^{-3}$ mbar while the yellow in 1000 mbar. The labels $t_1, \ldots, t_5$ relate the evolution of the electrical resistance in Fig. S1(a) with the sequence of T ramps and plateaux illustrated in Fig. S1(b). In all cases ramp speed was set to 0.1 K/s All the curves show a semiconducting behaviour, as expected for this film thickness. The absolute resistance value is affected by the oxygen pressure: in $10^{-3}$ mbar ($t_2$) the resistance increases, while in 1000 mbar ($t_4$) it decreases. This is a signature of hole-dominated transport, and is in-line with the back-gate measurements of Fig. 1 in the main text.

We note that the results of both back-gate measurements and $V_O$ doping experiments could be also interpreted as coming from a change in the carriers scattering times. In this scenario, the variation of $\sigma_0$ with the electrostatic gating could be related to the movement of the charge carriers towards defects at the $SrIrO_3/SrTiO_3$ interface, and the variations with $V_O$ could be related to the increase of point defects. However, our interpretation of an oxygen vacancies doping mechanism as a source of the observed variations in the electrical resistance is fully consistent with the results of the analysis presented in the main text, where we show that the electrical conductivity is dominated by holes. Furthermore, although both the capping layer and the substrate of $SrTiO_3$ could be affected by the $V_O$ doping, the observed response is not compatible with a doping of $SrTiO_3$ by oxygen vacancies, since it would result in an increased conductivity.

**Sec. 2. Sampling algorithm**

The mapping is performed using logarithmic sampling on a grid of $1000 \times 1000$ points. Initial boundaries are $n_\mathrm{e} \in [10^{20}, 10^{29}]\, 1/\mathrm{m}^3$ and $\mu_\mathrm{e} \in [10^{-1}, 10^3]\, \mathrm{cm}^2/\mathrm{Vs}$. These ranges are then narrowed around the valid combinations (see later) in order to increase the sampling resolution. The $(n_i, \mu_i)$ combinations compatible with the experimental data are calculated as follows:

1. A couple $(n_e, \mu_e)$ is taken in the probed domain and then the corresponding $(n_h, \mu_h)$ is calculated by:

$$n_h = \frac{(\sigma_0/e - n_e\mu_e)^2}{(\rho_H \sigma_0^2/e + n_e\mu_e^2)}, \quad \mu_h = \frac{(\rho_H \sigma_0^2/e + n_e\mu_e^2)}{(\sigma_0/e - n_e\mu_e)},$$

where $\rho_\mathrm{H}$ and $\sigma_0$ are experimental values. If the calculated $n_h$ or $\mu_h$ are negative the initial $(n_e, \mu_e)$ is discarded.

2. The Hall effect is calculated with both its full expression and the low-field linear approximation:

$$\rho_\mathrm{H}^{full} = \frac{1}{e} \frac{n_\mathrm{h}\mu_\mathrm{h}^2 - n_\mathrm{e}\mu_\mathrm{e}^2 + (n_\mathrm{h} - n_\mathrm{e})(\mu_\mathrm{h}\mu_\mathrm{e}B)^2}{\sigma_0^2/e^2 + (n_\mathrm{h} - n_\mathrm{e})^2(\mu_\mathrm{h}\mu_\mathrm{e}B)^2} \quad \rho_\mathrm{H}^{lin} = \frac{e}{\sigma_0^2}\left(n_\mathrm{h}\mu_\mathrm{h}^2 - n_\mathrm{e}\mu_\mathrm{e}^2\right)$$

The $(n_i, \mu_i)$ set is accepted only if the difference between $\rho_\mathrm{H}^{full}$ and $\rho_\mathrm{H}^{lin}$ on the $\pm 33\,\mathrm{T}$ range is less than 5 %, consistent with our experimental error in the measured magnetic field range. This condition is evaluated by calculating the difference between $\rho_\mathrm{H}^{full}$ and $\rho_\mathrm{H}^{lin}$ every 2 T and checking that nowhere it exceeds 5 %.

3. The expected Seebeck coefficients corresponding to the $(n_e, \mu_e)$ set are calculated using both Mott and Heikes formulas (see main text) and compared with the experimental values. If the results are within the acceptance ranges indicated in Table I of the main text, the $(n_i, \mu_i)$ set is accepted.

**Sec. 3. Analysis of experimental and calculated Magneto-Resistance**

From the results of Table II it is possible to calculate the expected cyclotron component of the magnetoresistance for a multiband system, that is given by:

$$\text{MR} = \frac{1}{e}\frac{n_h\mu_h + n_e\mu_e + (n_h\mu_e + n_e\mu_h)(\mu_e\mu_h B)^2}{\sigma_0^2/e^2 + (n_h - n_e)^2(\mu_h\mu_e B)^2}. \quad (1)$$

Fig. S2(a) shows the cyclotron component of the magneto-resistance (MR) calculated from the $(n_i, \mu_i)$ sets of Fig. 4. In order to compare the calculated MR with the experimental data, we consider the magneto-resistance measured on a thick sample of 30 uc. This is because thick samples are less affected by finite size-effects, such as quantum corrections, that become progressively more dominant for lower thicknesses[1]. The calculated cyclotron component is small compared to the low temperature experimental measurements presented in Fig. S2(b) and (c), which, on the other hand, show a linear dependence on the magnetic field. A better agreement between the calculated and experimental curves is found at intermediate temperatures, between 20 K and 120 K, where the field-dependence is parabolic. We interpret this as due to a small temperature-independent quadratic contribution (cyclotron component) that becomes negligible at low temperature if compared to other sources of magneto-resistance. This interpretation is supported by the fact that, as discussed in the main text, the experimental parameters determining the transport coefficient of the charge carriers are weakly dependent from temperature and, as a consequence of eq.1, the same behaviour is expected from the cyclotron component of the magneto-resistance, i.e. its magnitude is weakly affected by temperature.

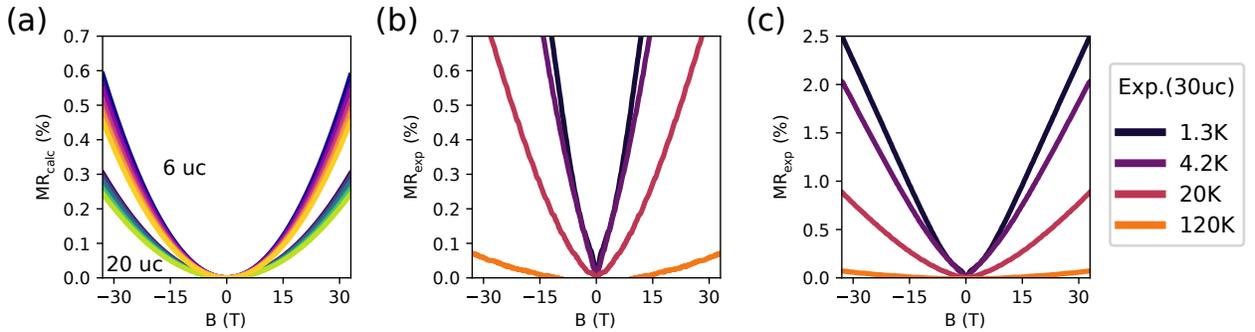

Figure S2. (a) Semi-classical magneto-resistance calculated from the $(n_i, \mu_i)$ sets from Fig. 4. (b) and (c) Experimental MR measured on a 30 uc film as a function of temperature temperatures. Plots show different magnification along the y axis for better comparison.

Within this picture, the observed linear trend should have a different origin.

Linear and non-saturating magneto-resistance at high field is similar to what is observed in several semimetallic compounds, as well as in the bulk form of SrIrO$_3$[2], and can be related to different mechanisms. Although it is typically due to the presence of Dirac cones[2–5], here this is unlikely because in SrIrO$_3$ deposited on top of SrTiO$_3$(100) Dirac cones are gapped[6]. Other possible sources of linear magneto-resistance in thin films are the formation of a ferromagnetic state[7], the presence of scattering centres[8] or spatial inhomogeneities in the transport parameters[9,10]. Linear magneto-resistance is also expected in finite-size samples of almost-compensated semimetals due to the interplay between bulk and edge electron-holes recombination[11]. Considering previous reports of diverging magnetic susceptibility at low temperatures, the possible presence of point defects and the semimetallic nature of SrIrO$_3$ thin films, several concurrent mechanisms are thus possible. Regardless of the specific source of linear magneto-resistance, we note that all these contributions get weaker when the temperature increases, making the cyclotron component more and more visible. This is in agreement with Fig. S2(b), where above 20 K, the measured MR assumes values close to the simulated semi-classical response of Fig. S2(a).


\end{document}